\documentclass[twocolumn,times]{aastex63}
\usepackage{graphicx}
\usepackage{makecell}
\usepackage{amsmath}
\usepackage{amsthm}
\usepackage{booktabs}
\usepackage{lineno}
\usepackage{float} 
\usepackage{threeparttable}
\usepackage{CJKutf8}
\usepackage[squaren]{SIunits}
\usepackage{xspace}
\usepackage{comment}
\usepackage{xcolor}
\usepackage{multirow}

\usepackage{array}

\setwatermarkfontsize{1.5in}

\shortauthors{Xiao ET AL.}
\begin{document}
\begin{CJK}{UTF8}{gbsn}

\title{Calibration of the Timing Performance of GECAM-C}
\correspondingauthor{Y.Q. Liu, K. Gong, Z.H. An}
\email{liuyaqing@ihep.ac.cn, gongk@ihep.ac.cn, anzh@ihep.ac.cn}

\author{Shuo Xiao}
\affil{School of Physics and Electronic Science, Guizhou Normal University, Guiyang 550001, People’s Republic of China;}
\affil{Guizhou Provincial Key Laboratory of Radio Astronomy and Data Processing, Guizhou Normal University, Guiyang 550001, People’s Republic of China}

\author{Ya-Qing Liu*}
\affil{Key Laboratory of Particle Astrophysics, Institute of High Energy Physics, Chinese Academy of Sciences, Beijing 100049, China}

\author{Ke Gong*}
\affil{Key Laboratory of Particle Astrophysics, Institute of High Energy Physics, Chinese Academy of Sciences, Beijing 100049, China}

\author{Zheng-Hua An*}
\affil{Key Laboratory of Particle Astrophysics, Institute of High Energy Physics, Chinese Academy of Sciences, Beijing 100049, China}

\author{Shao-Lin Xiong}
\affil{Key Laboratory of Particle Astrophysics, Institute of High Energy Physics, Chinese Academy of Sciences, Beijing 100049, China}

\author{Xin-Qiao Li}
\affil{Key Laboratory of Particle Astrophysics, Institute of High Energy Physics, Chinese Academy of Sciences, Beijing 100049, China}

\author{Xiang-Yang Wen}
\affil{Key Laboratory of Particle Astrophysics, Institute of High Energy Physics, Chinese Academy of Sciences, Beijing 100049, China}

\author{Wen-Xi Peng}
\affil{Key Laboratory of Particle Astrophysics, Institute of High Energy Physics, Chinese Academy of Sciences, Beijing 100049, China}

\author{Da-Li Zhang}
\affil{Key Laboratory of Particle Astrophysics, Institute of High Energy Physics, Chinese Academy of Sciences, Beijing 100049, China}

\author{You-Li Tuo}
\affil{Institut für Astronomie und Astrophysik, University of Tübingen, Sand 1, 72076 Tübingen, Germany}

\author{Shi-Jie Zheng}
\affil{Key Laboratory of Particle Astrophysics, Institute of High Energy Physics, Chinese Academy of Sciences, Beijing 100049, China}

\author{Li-Ming Song}
\affil{Key Laboratory of Particle Astrophysics, Institute of High Energy Physics, Chinese Academy of Sciences, Beijing 100049, China}

\author{Ping Wang}
\affil{Key Laboratory of Particle Astrophysics, Institute of High Energy Physics, Chinese Academy of Sciences, Beijing 100049, China}

\author{Xiao-Yun Zhao}
\affil{Key Laboratory of Particle Astrophysics, Institute of High Energy Physics, Chinese Academy of Sciences, Beijing 100049, China}

\author{Yue Huang}
\affil{Key Laboratory of Particle Astrophysics, Institute of High Energy Physics, Chinese Academy of Sciences, Beijing 100049, China}

\author{Xiang Ma}
\affil{Key Laboratory of Particle Astrophysics, Institute of High Energy Physics, Chinese Academy of Sciences, Beijing 100049, China}

\author{Xiao-Jing Liu}
\affil{Key Laboratory of Particle Astrophysics, Institute of High Energy Physics, Chinese Academy of Sciences, Beijing 100049, China}

\author{Rui Qiao}
\affil{Key Laboratory of Particle Astrophysics, Institute of High Energy Physics, Chinese Academy of Sciences, Beijing 100049, China}

\author{Yan-Bing Xu}
\affil{Key Laboratory of Particle Astrophysics, Institute of High Energy Physics, Chinese Academy of Sciences, Beijing 100049, China}

\author{Sheng Yang}
\affil{Key Laboratory of Particle Astrophysics, Institute of High Energy Physics, Chinese Academy of Sciences, Beijing 100049, China}

\author{Fan Zhang}
\affil{Key Laboratory of Particle Astrophysics, Institute of High Energy Physics, Chinese Academy of Sciences, Beijing 100049, China}

\author{Yue Wang}
\affil{Key Laboratory of Particle Astrophysics, Institute of High Energy Physics, Chinese Academy of Sciences, Beijing 100049, China}
\affil{University of Chinese Academy of Sciences, Chinese Academy of Sciences, Beijing 100049, China}

\author{Yan-Qiu Zhang}
\affil{Key Laboratory of Particle Astrophysics, Institute of High Energy Physics, Chinese Academy of Sciences, Beijing 100049, China}
\affil{University of Chinese Academy of Sciences, Chinese Academy of Sciences, Beijing 100049, China}

\author{Wang-Chen Xue}
\affil{Key Laboratory of Particle Astrophysics, Institute of High Energy Physics, Chinese Academy of Sciences, Beijing 100049, China}
\affil{University of Chinese Academy of Sciences, Chinese Academy of Sciences, Beijing 100049, China}

\author{Jia-Cong Liu}
\affil{Key Laboratory of Particle Astrophysics, Institute of High Energy Physics, Chinese Academy of Sciences, Beijing 100049, China}
\affil{University of Chinese Academy of Sciences, Chinese Academy of Sciences, Beijing 100049, China}

\author{Chao Zheng}
\affil{Key Laboratory of Particle Astrophysics, Institute of High Energy Physics, Chinese Academy of Sciences, Beijing 100049, China}
\affil{University of Chinese Academy of Sciences, Chinese Academy of Sciences, Beijing 100049, China}

\author{Chen-Wei Wang}
\affil{Key Laboratory of Particle Astrophysics, Institute of High Energy Physics, Chinese Academy of Sciences, Beijing 100049, China}
\affil{University of Chinese Academy of Sciences, Chinese Academy of Sciences, Beijing 100049, China}

\author{Wen-Jun Tan}
\affil{Key Laboratory of Particle Astrophysics, Institute of High Energy Physics, Chinese Academy of Sciences, Beijing 100049, China}
\affil{University of Chinese Academy of Sciences, Chinese Academy of Sciences, Beijing 100049, China}

\author{Ce Cai}
\affil{College of Physics, Hebei Normal University, 20 South Erhuan Road, Shijiazhuang, 050024, China}

\author{Qi-Bin Yi}
\affil{Key Laboratory of Particle Astrophysics, Institute of High Energy Physics, Chinese Academy of Sciences, Beijing 100049, China}
\affil{Key Laboratory of Stellar and Interstellar Physics and Department of Physics, Xiangtan University, 411105 Xiangtan, Hunan Province, China}

\author{Peng Zhang}
\affil{Key Laboratory of Particle Astrophysics, Institute of High Energy Physics, Chinese Academy of Sciences, Beijing 100049, China}
\affil{College of Electronic and Information Engineering, Tongji University, Shanghai 201804, China}

\author{Xi-Hong Luo}
\affil{School of Physics and Electronic Science, Guizhou Normal University, Guiyang 550001, People’s Republic of China;}
\affil{Guizhou Provincial Key Laboratory of Radio Astronomy and Data Processing, Guizhou Normal University, Guiyang 550001, People’s Republic of China}

\author{Jiao-Jiao Yang}
\affil{School of Physics and Electronic Science, Guizhou Normal University, Guiyang 550001, People’s Republic of China;}
\affil{Guizhou Provincial Key Laboratory of Radio Astronomy and Data Processing, Guizhou Normal University, Guiyang 550001, People’s Republic of China}

\author{Qi-Jun Zhi}
\affil{School of Physics and Electronic Science, Guizhou Normal University, Guiyang 550001, People’s Republic of China;}
\affil{Guizhou Provincial Key Laboratory of Radio Astronomy and Data Processing, Guizhou Normal University, Guiyang 550001, People’s Republic of China}

\author{Ai-Jun Dong}
\affil{School of Physics and Electronic Science, Guizhou Normal University, Guiyang 550001, People’s Republic of China;}
\affil{Guizhou Provincial Key Laboratory of Radio Astronomy and Data Processing, Guizhou Normal University, Guiyang 550001, People’s Republic of China}

\author{Shi-Jun Dang}
\affil{School of Physics and Electronic Science, Guizhou Normal University, Guiyang 550001, People’s Republic of China;}
\affil{Guizhou Provincial Key Laboratory of Radio Astronomy and Data Processing, Guizhou Normal University, Guiyang 550001, People’s Republic of China}

\author{Lun-Hua Shang}
\affil{School of Physics and Electronic Science, Guizhou Normal University, Guiyang 550001, People’s Republic of China;}
\affil{Guizhou Provincial Key Laboratory of Radio Astronomy and Data Processing, Guizhou Normal University, Guiyang 550001, People’s Republic of China}

\author{Shuang-Nan Zhang}
\affil{Key Laboratory of Particle Astrophysics, Institute of High Energy Physics, Chinese Academy of Sciences, Beijing 100049, China}
\affil{University of Chinese Academy of Sciences, Chinese Academy of Sciences, Beijing 100049, China}

\begin{abstract}
As a new member of the Gravitational wave high-energy Electromagnetic Counterpart All-sky Monitor (GECAM) after GECAM-A and GECAM-B, GECAM-C (originally called HEBS), which was launched on board the SATech-01 satellite on July 27, 2022, aims to monitor and localize X-ray and gamma-ray transients from $\sim$ 6 keV to 6 MeV. GECAM-C utilizes a similar design to GECAM but operates in a more complex orbital environment. In this work, we utilize the secondary particles simultaneously produced by the cosmic-ray events on orbit and recorded by multiple detectors, to calibrate the relative timing accuracy between all detectors of GECAM-C. We find the result is 0.1 $\mu \rm s$, which is the highest time resolution among all GRB detectors ever flown and very helpful in timing analyses such as minimum variable timescale and spectral lags, as well as in time delay localization. Besides, we calibrate the absolute time accuracy using the one-year Crab pulsar data observed by GECAM-C and Fermi/GBM, as well as GECAM-C and GECAM-B. The results are $2.02\pm 2.26\ \mu \rm s$ and $5.82\pm 3.59\ \mu \rm s$, respectively. Finally, we investigate the spectral lag between the different energy bands of Crab pulsar observed by GECAM and GBM, which is $\sim -0.2\ {\rm \mu s\ keV^{-1}}$.

\end{abstract}


\section{Introduction}
Since the launch on December 10, 2020, the GECAM mission \citep{xiong2020gecam} has observed hundreds of GRBs including the Brightest-Of-All-Time (BOAT) GRB 221009A \citep{an2023insight} and the second brightest GRB 230307A \citep{xiong2023grb}, as well as X-ray bursts from magnetar such as SGR J1935+2154. Thanks to the excellent energy resolution ($\sim$ 16\% at 59.5 keV), anti-saturation design at high count rates ($<$ 4$\times 10^5$ counts per second) (\citealp{zhang2019energy}; \citealp{chen2021design}; \citealp{li2021inflight}) and the highest time resolution (0.1 $\mu \rm s$) among all GRB satellites ever flown \citep{xiao2022ground}, GECAM has advantages in energy spectral and timing analysis \citep{xiao2022energetic}, such as spectral lag \citep{xiao2022robust,xiao2023discovery}, minimum variable timescale \citep{xiao2023minimum} and quasiperiodic oscillation \citep{xiao2023individual}, as well as time delay localization for burst \citep{xiao2021enhanced} and pulsar or magnetar navigation experiment \citep{luo2023pulsar}.

\begin{figure*}
\centering
\begin{minipage}[t]{0.99\textwidth}
\centering
\includegraphics[width=\columnwidth]{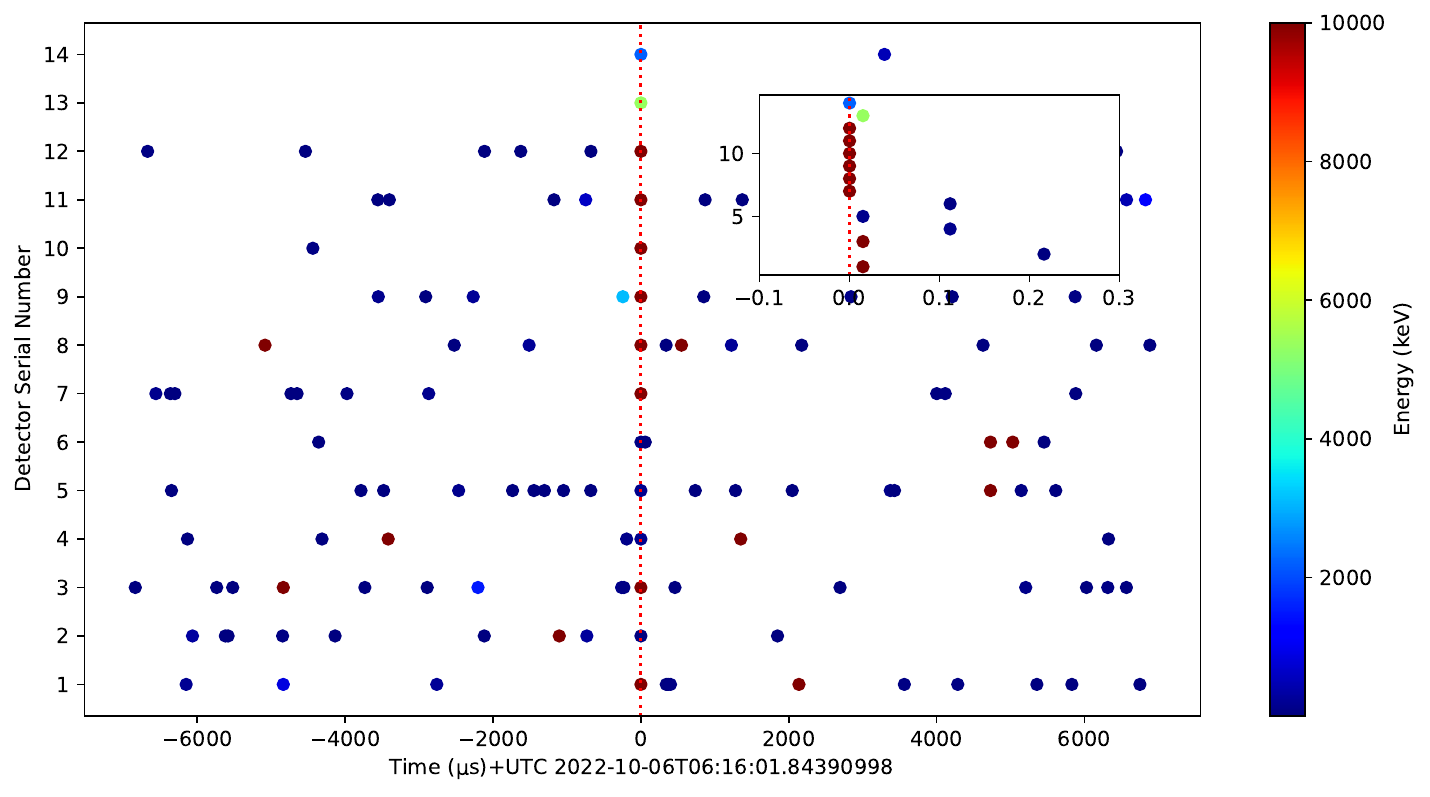}
\end{minipage}
\caption{An illustration of the secondary particles produced by cosmic rays hitting the satellite and observed by all 14 detectors on board GECAM-C simultaneously at UTC 2022-10-06T06:16:01.84390998 (red dotted line). The vertical coordinate represents the detector serial number, from 1 to 12 represent GRD, whereas 13 and 14 represent CPD. The different colors represent different energies, note that there are 8 GRDs recorded with energies higher than 10,000 keV due to the very high energy of the cosmic ray. Note their time difference is less than 0.3 $\rm \micro\second$.}
\label{a_cos}
\end{figure*}

To achieve all-sky monitoring, joint data analysis, localization of bursts, as well as the study of self-organized criticality \citep{bak1987self} for a more comprehensive sample, especially since GECAM-A is currently not stable for operation, GECAM-C (i.e. the High Energy Burst Searcher, HEBS) is designed as a new member of GECAM on board the SATech-01 satellite \footnote{https://www.globaltimes.cn/page/202301/1283692.shtml} launched on July 27, 2022. GECAM-C \citep{ZHANG2023168586} consists of two dome modules which are installed on the top and bottom side of the satellite (see Figure 1 in \cite{zheng2023ground}), each dome module has one data acquisition board (DAQ) to read out data from six Gamma-Ray Detectors (GRDs) and one Charged Particle
Detector (CPD). An encapsulated crystal box placed NaI(Tl) or LaBr$_3$ (Ce/Ce+Sr) crystal and a read out by a SiPM array are utilized in each GRD design, the preamplifier of each GRD has two channels (high gain and low gain), except for GRD06 and GRD12 detectors, which are readout by a single channel. Therefore, the detectors installation positions and detectors design are not the same as in GECAM-A or GECAM-B.
The detection range of GECAM-C is about
6 keV–6 MeV with an energy resolution of 18\% at 59.5 keV \citep{zheng2023ground}. The dead time for a normal event (i.e. that with energy within the dynamic range of GRD or CPD readout) is 4 $\mu \rm s$ and 4.8 $\mu \rm s$ for GRD and CPD, respectively.

The time system of GECAM-C is the same as that of GECAM-B (see \cite{xiao2022ground} for details), except that the former has only two DAQs and the time accuracy of GNSS receiver and hardware delay is about +/-400 nanoseconds (3$\sigma$). The timing performance, including relative time accuracy between all detectors of GECAM-C and absolute time accuracy, is one of the core metrics of GRB satellites. In the previous calibration  of the timing performance of GECAM-B, we proposed using cosmic ray events for relative time accuracy calibration and obtain an accuracy of 0.12 $\mu \rm s$ (1$\sigma$), and the data of Crab pulsar observed by GECAM-B and {\it Fermi}/GBM from 2020 December
to 2021 August is utilized to calibrate the absolute time accuracy and obtain an accuracy of $3.06 \pm 6.04\ \mu \rm s$ (1$\sigma$). Additionally, we introduced a new method for absolute time calibration using bursts from a magnetar, such as SGR J1935+2154 \citep{xiao2022ground}. Although GECAM and GBM have similar responses and been chosen with the same energy range, we may be unable to completely avoid systematic error due to the spectral lags between different energy bands in Crab pulsar with $t_{\rm lag}(E)=-0.27(E/{\rm keV})+C$ ($\mu \rm s$) (i.e. low-energy photons arrive before high-energy photons.) \citep{molkov2009absolute,tuo2022orbit}. Fortunately, for the very similarly designed GECAM-B and GECAM-C, we have the valuable opportunity to perform time calibration through the Crab pulsar data observed by them.

In this work, the calibration of the relative time accuracy of 12 GRDs and 2 CPDs onboard GECAM-C using cosmic rays observed on orbit is presented in Section 2. We perform the absolute time calibration in Section 3 using the data of Crab pulsar observed by GECAM-C and {\it Fermi}/GBM, as well as GECAM-C and GECAM-B. A summary is given in Section 4.

\section{RELATIVE TIME CALIBRATION}
\begin{figure*}
\centering
\begin{minipage}[t]{0.49\textwidth}
\centering
\includegraphics[width=\columnwidth]{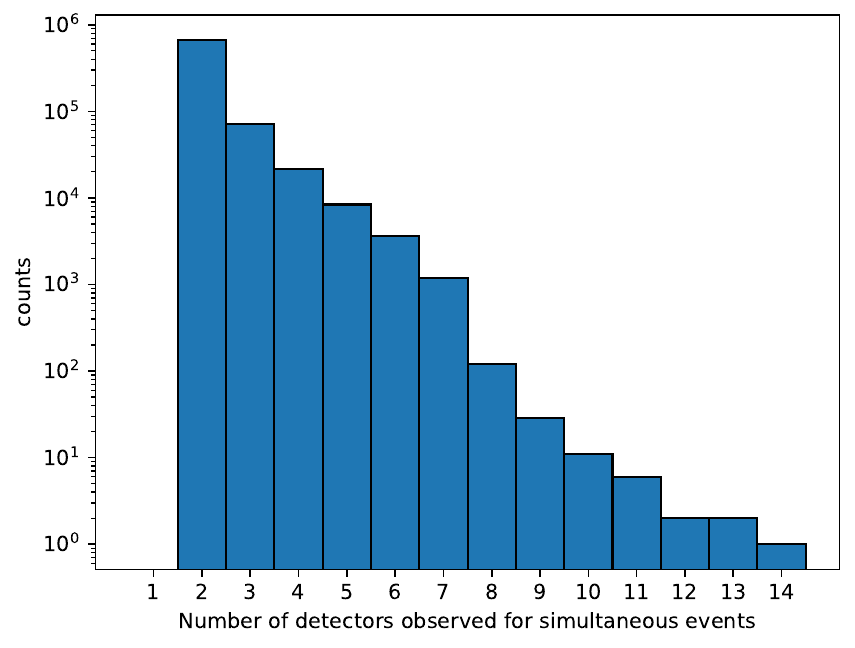}
\end{minipage}
\begin{minipage}[t]{0.49\textwidth}
\centering
\includegraphics[width=\columnwidth]{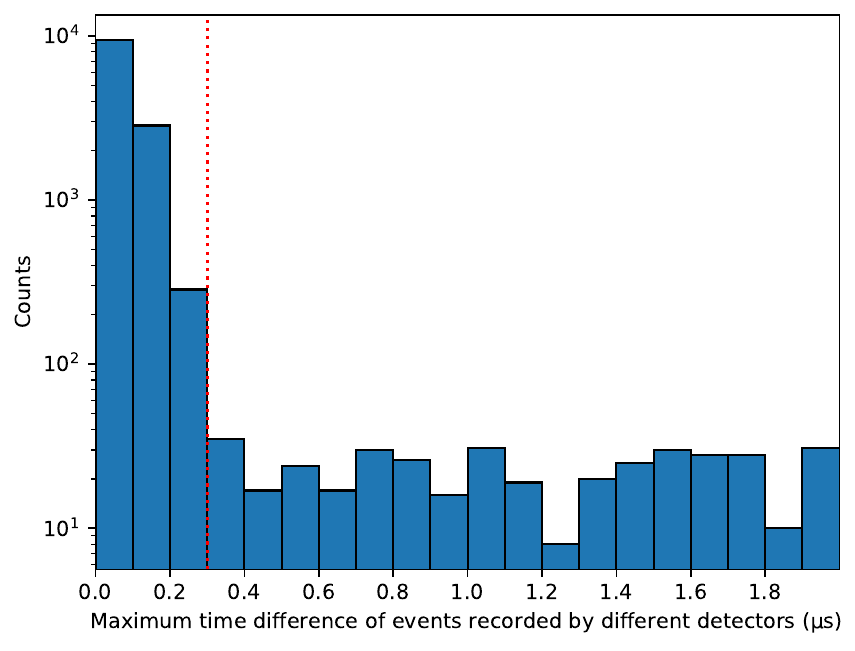}
\end{minipage}
\caption{Left panel: the distribution of the number of detectors marked as observed simultaneous events in an hour (from UTC 2022-10-06T06:00:00 to 2022-10-06T06:59:59), the number of detectors with more than seven observing cosmic ray events is suddenly much fewer since each group of 7 detectors on board GECAM-C are mounted at each side of the satellite and pointing in opposite directions relative to the other group (i.e. higher-energy cosmic rays are needed to be observed by more than seven detectors). Right panel: the distribution of the maximum time difference for 13353 cosmic ray events observed by multiple detectors, most of them (97.0\%) are distributed within less than 0.3 $\rm \micro\second$ and are approximately uniformly distributed after more than 0.3 $\rm \micro\second$, which is produced by unrelated individual events.}\label{max_delta_1}
\end{figure*}

When a cosmic ray (mostly proton) hits a satellite, it may produce many secondary particles through hadron shower or other processes observed by multiple detectors almost simultaneously. Figure~\ref{a_cos} shows a time vs. detector serial number scatter plot for a very high energy cosmic ray event, whose secondary particles are observed by all 14 detectors at UTC 2022-10-06T06:16:01.84390998, note that there are 8 GRDs recorded with energies higher than 10,000 keV. All 14 detectors detected within 0.3 $\mu \rm s$ of each other, thus, similarly to GECAM-B, we initially set up that if more than 1 detector observes events on a time window of +-0.3 $\rm \micro\second$, they are considered simultaneous events produced by a cosmic ray event. 
The left panel in Figure~\ref{max_delta_1} shows the distribution of the number of detectors observed cosmic ray events (i.e. marked as simultaneous events) in an hour (from UTC 2022-10-06T06:00:00 to 2022-10-06T06:59:59). It is worth noting here that unlike GECAM-A or GECAM-B where all detectors are installed on a dome module, here each group of the 7 detectors of GECAM-C in each dome at each side of the satellite point in opposite directions relative to he other group, thus the number of detectors with more than 7 observing the same cosmic ray events suddenly declines. 

However, if the relative time accuracy exceeds 0.3 $\rm \micro\second$, there might be events observed by some detectors not marked as simultaneous events according to our set threshold. To check this, we analyze 13353 cosmic ray events in which at least five detectors are marked as observing simultaneous events during this hour, to avoid the coincident by unrelated individual events as much as possible. Then we search all events recorded by all 14 detectors in a time window of $\pm$1 $\rm \micro\second$ before and after the recording time of the simultaneous events, and calculating the maximum time difference among them. The distribution of the maximum time difference is shown in the right panel of Figure~\ref{max_delta_1}, where we find that most of them (97.0\%) are distributed within less than 0.3 $\rm \micro\second$ and the others are approximately uniformly distributed after more than 0.3 $\rm \micro\second$; the latter is produced by not simultaneous events (i.e. coincident by unrelated individual events) and indicates that this result is not affected by the time window (i.e. $\pm$1 $\rm \micro\second$) we set for this analysis. Besides, about 76\% of the maximum time difference is less than or equal to 0.1 $\rm \micro\second$, thus the relative time accuracy between the detectors onboard GECAM-C is $\sim$ 0.1 $\rm \micro\second$. 

\begin{figure}
\centering
\includegraphics[width=\columnwidth]{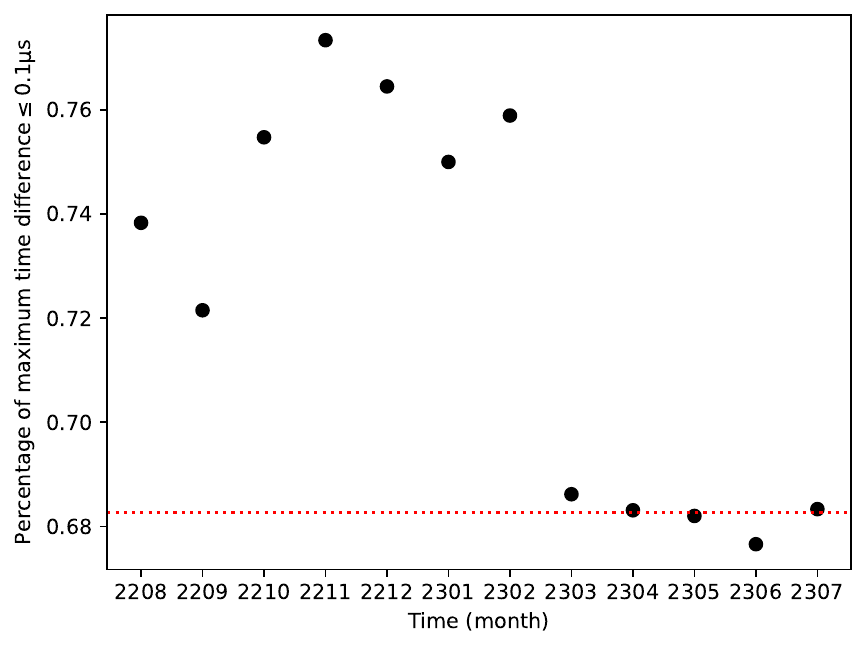}
\caption{The percentage of maximum time differences of less than or equal to 0.1 $\rm \micro\second$ for simultaneous events observed by multiple detectors in each month from 2022 August to 2023 July (i.e. vertical coordinate). The red dotted line represents the 1 $\sigma$ in the Gaussian distribution. Therefore, the time resolution of GECAM-C is better than 0.1 $\mu \rm s$ (1 $\sigma$) due to the fact that the percentage in most months greater than 68\%.}
\label{max_delta_month}
\end{figure}

Furthermore, we employ the procedure above to analyze the cosmic ray events in each month from 2022 August to 2023 July to test whether the relative time accuracy between the detectors changes over time. As shown in Figure \ref{max_delta_month}, the percentages of the maximum time differences of less than or equal to 0.1 $\rm \micro\second$ for simultaneous events observed by multiple detectors obtained in most months are greater than 68\%, which demonstrates the stability of the time resolution. In addition, we note a tendency for the percentage to first increase and then decrease with time, for which we have not yet identified a reason, but have ruled out temperature and attitude changes. However, this does not affect the conclusion of 0.1 $\rm \micro\second$ time resolution.

\begin{figure}
\centering
\includegraphics[width=\columnwidth]{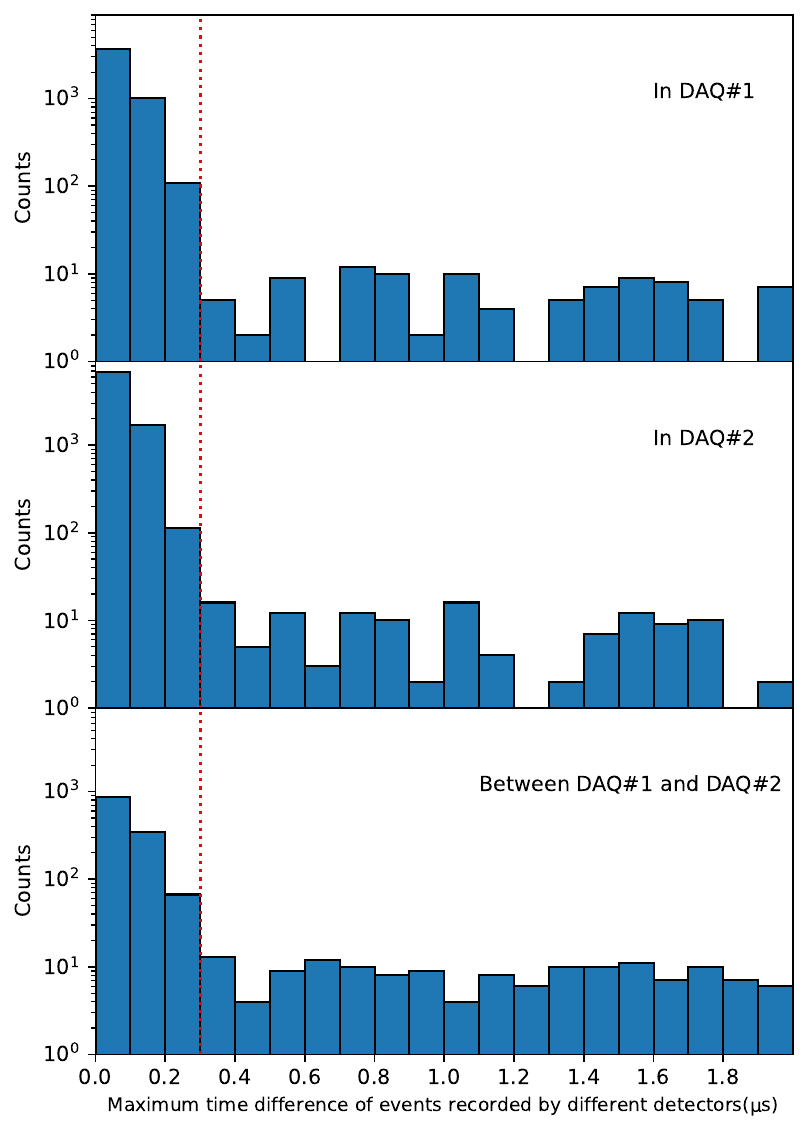}
\caption{The distribution of the maximum time difference for 13353 cosmic ray events observed by multiple detectors in DAQ\#1 (top panel), DAQ\#2 (middle panel), as well as DAQ\#1 and DAQ\#2 (i.e. both DAQs need at least one detector detects the event) (bottom panel), respectively. 98.2\%, 98.6\% and 92.3\% of the events are distributed within less than 0.3 $\rm \micro\second$, respectively. Besides, there are 98.6\%, 99.0\% and 96.2\% of the events with maximum time difference of less than 0.2 $\rm \micro\second$ in the events of less than 0.3 $\rm \micro\second$, respectively. Therefore, the time accuracy between detectors in two DAQs is slightly worse than that of detectors in the same DAQ.}
\label{time_daqs}
\end{figure}

We also respectively investigated the time accuracy between detectors in the same DAQ (i.e. DAQ\#1 or DAQ\#2), as well as between detectors in different DAQs. As shown in Figure \ref{time_daqs}, there are 98.2\%, 98.6\% and 92.3\% of the events with maximum time difference of less than 0.3 $\rm \micro\second$, respectively. Besides, there are 98.6\%, 99.0\% and 96.2\% of the events with maximum time difference of less than 0.2 $\rm \micro\second$ in the events of less than 0.3 $\rm \micro\second$, respectively. Therefore, the time accuracy between detectors in two DAQs is slightly worse than that of detectors in the same DAQ. It's worth noting here that both DAQs observed significantly fewer cosmic ray events than a single DAQ due to they are mounted at each side of the satellite and pointing in opposite directions relative to the other group.

\begin{figure*}
\centering
\begin{minipage}[t]{0.99\textwidth}
\centering
\includegraphics[width=\columnwidth]{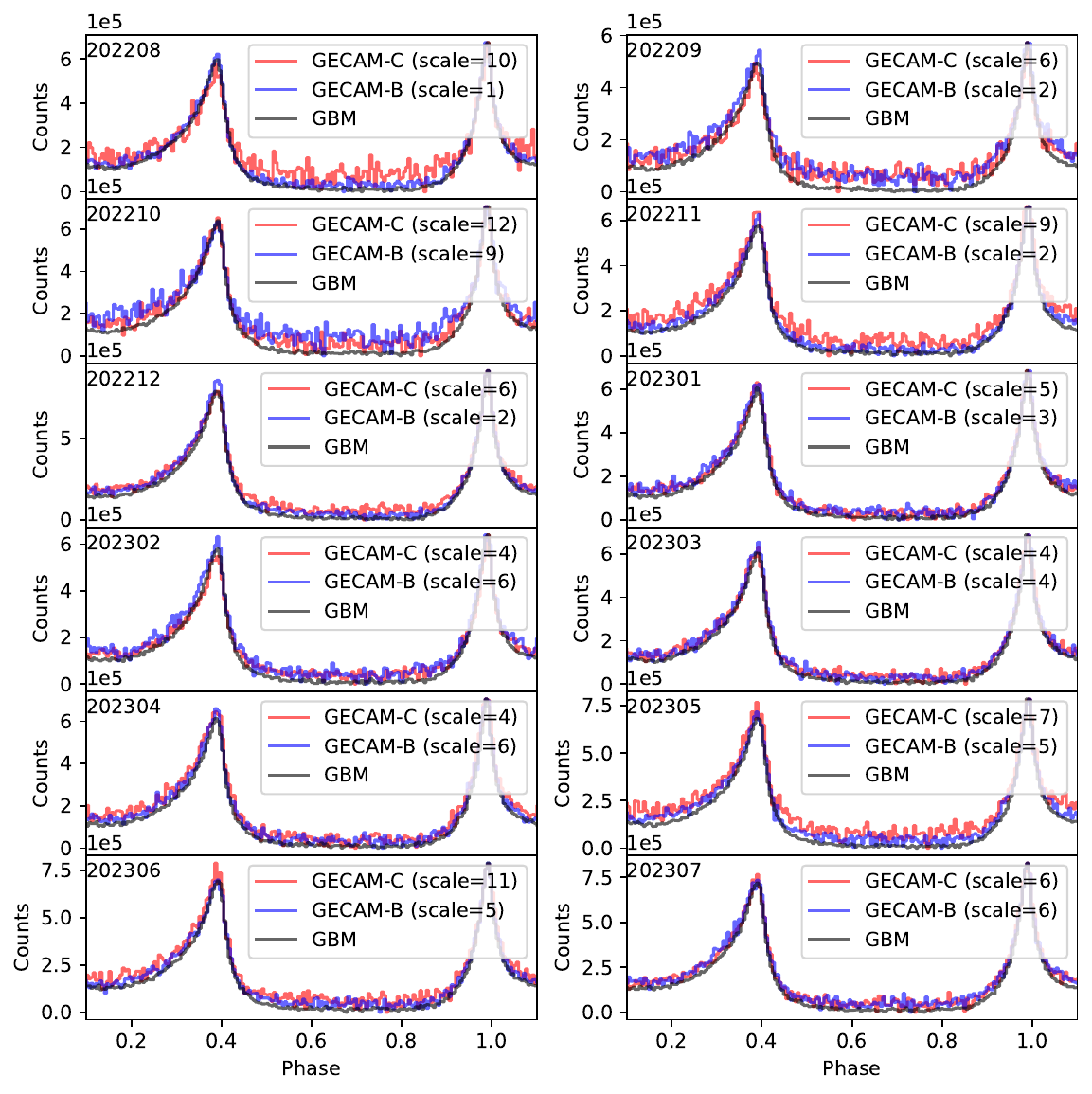}
\end{minipage}
\caption{The Crab pulse profiles in the 20–500 keV energy band observed GECAM-C (red), GECAM-B (blue) and GBM (black) in each month from August 2022 to July 2023. Scale represents the coefficient by which its vertical coordinate is multiplied, and we subtract the lowest value of the pulse profile, respectively.}
\label{crab_pulse}
\end{figure*}

\begin{figure}
\centering
\includegraphics[width=\columnwidth]{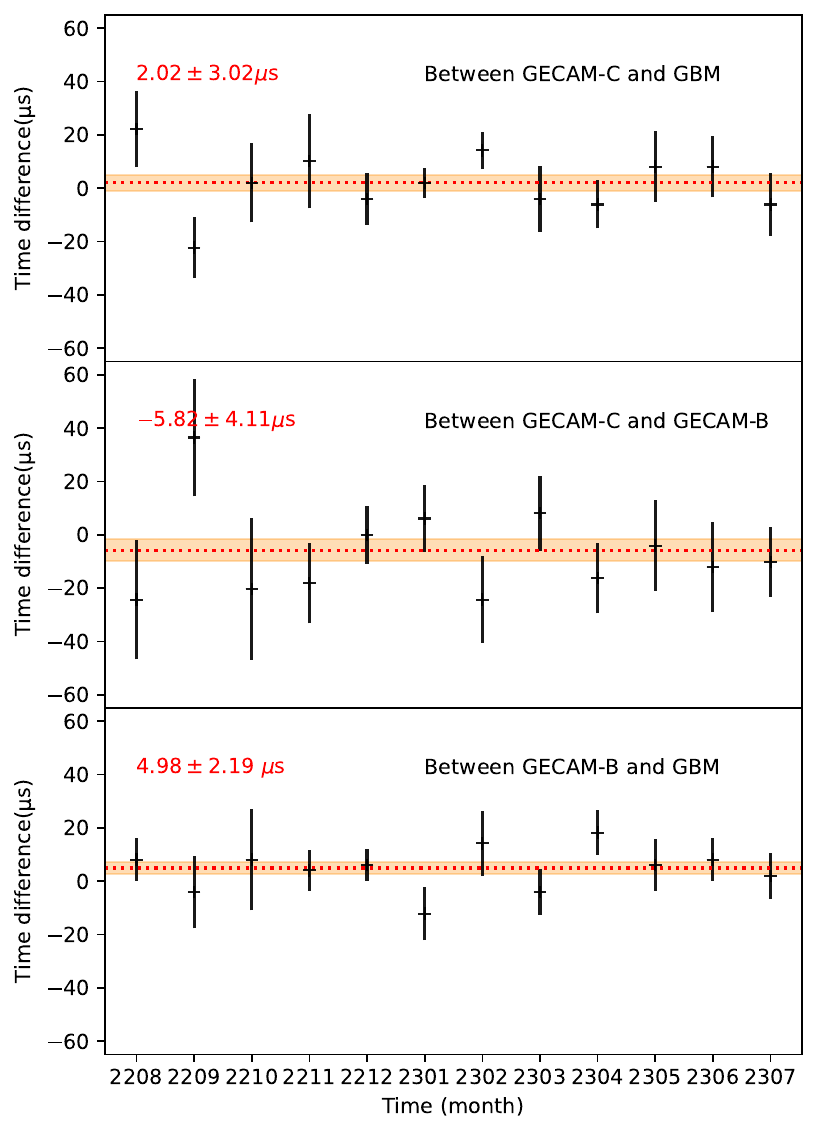}
\caption{The time difference of observed Crab pulse profiles between GECAM-C and GBM, between GECAM-C and GECAM-B, as well as between GECAM-B and GBM in different months, respectively. The orange shaded regions are the 1$\sigma$ ranges (i.e. 2.02$\pm$ 3.02 $\rm \micro\second$, -5.82$\pm$ 4.11 $\rm \micro\second$ and 4.98$\pm$ 2.19 $\rm \micro\second$) obtained by fitting. The reduced-$\chi^2$ is 1.05, 0.94 and 0.86, respectively, which indicates their well fit goodness of fit.}
\label{errors}
\end{figure}

\begin{figure*}
\centering
\begin{minipage}[t]{0.49\textwidth}
\centering
\includegraphics[width=\columnwidth]{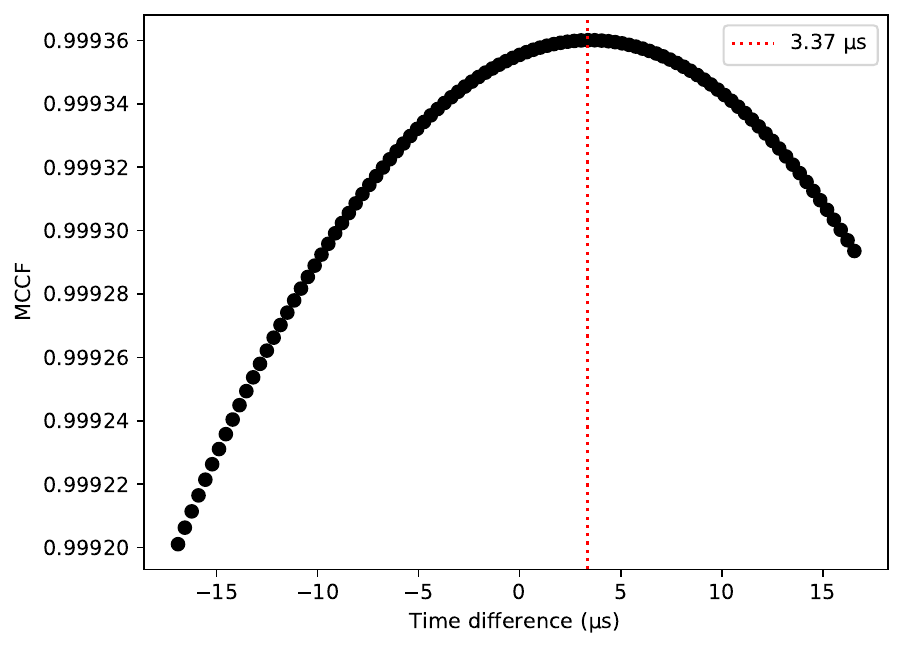}
\end{minipage}
\begin{minipage}[t]{0.49\textwidth}
\centering
\includegraphics[width=\columnwidth]{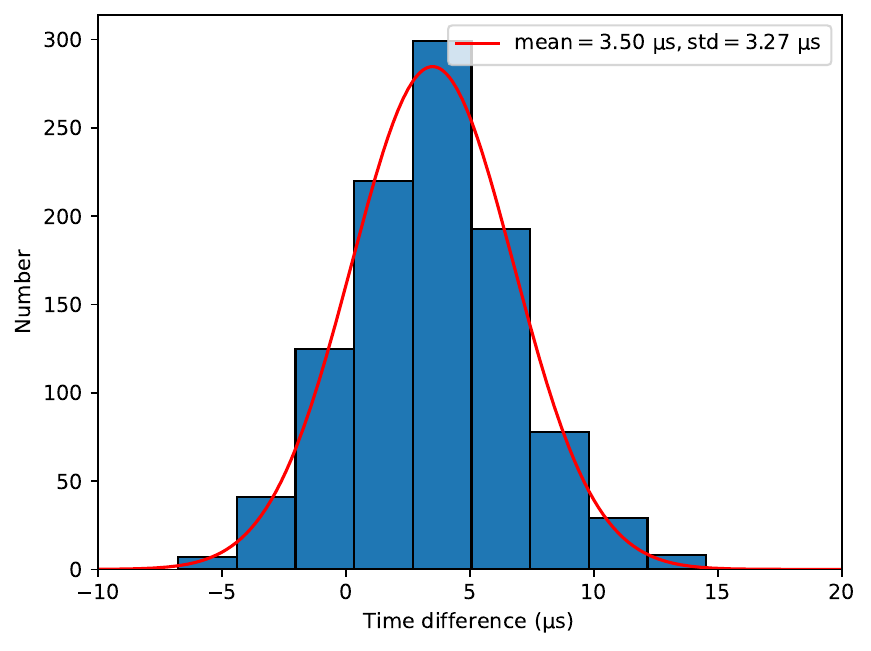}
\end{minipage}
\caption{Left panel: the black dots are the MCCF versus time difference of the between GECAM-C and GBM observations of Crab pulsar profiles in one year, the red dotted line represents the best time difference. Right panel: distribution of time differences obtained by MC, the red line is obtained by fitting a Gaussian function. Therefore, the time difference is $3.37\pm3.27\ \mu\second$, which is consistent with the result obtained by fitting each month (i.e. $2.02\pm3.02\ \mu\second$) within the error.}\label{MCCF_MC}
\end{figure*}

\begin{figure}
\centering
\includegraphics[width=\columnwidth]{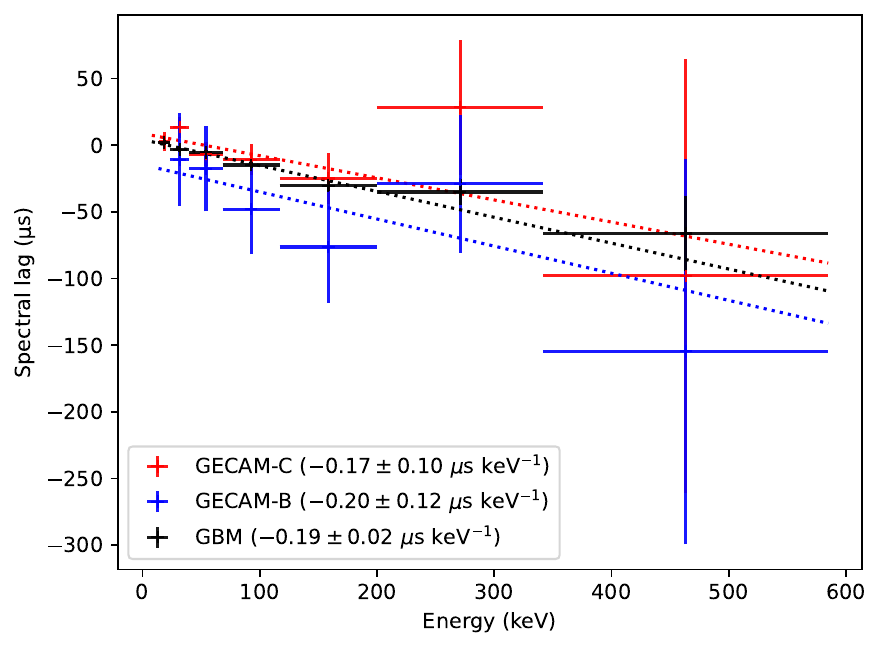}
\caption{Spectral lags of Crab profiles at different energies observed by GECAM-C (red), GECAM-B (blue) and GBM (black), respectively. Note that for GECAM-C and GBM the lag is compared to 8-14 keV and for GECAM-B the lag is compared to 14-23 keV. The dotted lines represent the results obtained by fitting them with a linear function, respectively.}
\label{lag_ee}
\end{figure}

\section{ABSOLUTE TIME CALIBRATION}
\subsection{Between GECAM-C and Fermi/GBM}
Crab pulsar (PSR J0534+2200, RA = $\rm 05^{h}34^{m}31.972^{s}$, Dec = $\rm 22 ^{\circ} 0'52".07$) \citep{lyne199323} is widely used in previous absolute time accuracy calibrations of almost all astronomical satellites (\citealp{2003A&A...411L..31K}; \citealp{rots2004absolute}; \citealp{kirsch2004timing}; \citealp{edwards2006tempo2}; \citealp{molkov2009absolute}; \citealp{terada2008orbit}; \citealp{cusumano2012timing}; \citealp{basu2018timing}; \citealp{bachetti2021timing}; \citealp{tuo2022orbit}; \citealp{xiao2022ground}) due to its relatively high brightness and stable evolutionary period. Since Fermi/GBM has the time accuracy of about 10 $\rm \micro\second$ and the time resolution of 2 $\rm \micro\second$ (\citealp{2009ApJ...702..791M}; \citealp{paciesas2012fermi}), as well as an energy response of the NaI detector similar to that of GECAM-C, we utilize the Crab pulsar data observed by GECAM-C and GBM from August 1, 2022 to July 31, 2023 for absolute time calibration. Besides, to improve the statistics, only those detectors on board GECAM-C or GBM with incidence angles less than 70 degrees from the Crab pulsar and only the time intervals when the Crab pulsar is not blocked by the Earth are selected, as well as counts in the deposition energy range of 20-500 keV (the energy responses of GBM and GECAM are similar in this energy range) are used to analysis. 

To calculate the phase in Crab pulsar profile of each event observed GECAM-C and GBM, the arrive times of them are corrected to the solar system barycenter (DE200) using the $\rm package$ $tat-pulsar$ \footnote{https://github.com/tuoyl/tat-pulsar} developed by us \citep{tuo2022orbit}, then folded with the radio ephemeris \footnote{http://www.jb.man.ac.uk/~pulsar/crab.html} \citep{lyne199323}, respectively. We then utilize the Li-CCF method \citep{li2004timescale,xiao2021enhanced}, which can obtain a more accurate result using as much information in the data as possible, to calculate the time difference between the Crab pulse profiles observed by GECAM-C and GBM, respectively, which is taken as the GECAM-C absolute time accuracy; the detailed procedures can be found in our previous work \citep{xiao2022ground}.

Figure~\ref{crab_pulse} shows the pulse profiles observed by GECAM-C, GECAM-B, and Fermi/GBM in each month from August 2022 through July 2023, which have very similar profiles. We calculate the time difference between the pulsar profiles observed by GECAM-C and GBM in each month using Li-CCF and the results are shown in Table \ref{allresults}. Since the results in different months are roughly consistent within the errors (see the top panel in Figure \ref{errors}), we obtain the absolute time accuracy by fitting with a constant, which is 2.02$\pm$ 3.02 $\rm \micro\second$ (reduced-$\chi^2$=1.05). Note that if we do not take into account the error introduced by the time resolutions of GBM (i.e. 2 $\rm \micro\second$) and GECAM-C (i.e. 0.1 $\rm \micro\second$), the statistical uncertainty on the absolute time accuracy is $\sqrt{3.02^2-2^2-0.1^2}=2.26\ \rm \micro\second$, i.e., the final absolute time accuracy is 2.02$\pm$ 2.26 $\rm \micro\second$, which is consistent with 0 within 0.9 $\sigma$ error.

In addition, we directly calculated the time difference between the Crab pulsar profiles in one year observed by GECAM-C and GBM. The left panel in Figure \ref{MCCF_MC} shows the MCCF versus time difference, the time difference corresponding to the maximum value of MCCF is the best value (i.e. 3.37 $\rm \mu\second$). The right panel in Figure \ref{MCCF_MC} shows the distribution of time differences obtained by a Monte Carlo (MC) simulation, since its approximation follows a Gaussian distribution, the standard deviation can be taken as the 1 $\sigma$ error. Therefore, the result is $3.37\pm3.27\ \mu\second$, which is consistent with the result obtained by fitting each month (i.e. $2.02\pm3.02\ \mu\second$) within the error.

\subsection{Between GECAM-B and Fermi/GBM}
Meanwhile, we can also further calibrate the absolute time accuracy of GECAM-B. The time difference between the pulsar profiles observed by GECAM-C and GBM in each month. The results are shown in Table \ref{allresults} and the final absolute time accuracy is 4.98$\pm$ 2.19 $\rm \micro\second$ (reduced-$\chi^2$=0.94) through fitting (see the middle panel in Figure \ref{errors}). If we do not take into account the error introduced by the time resolution of GBM (i.e. 2 $\rm \micro\second$) and GECAM-B (i.e. 0.1 $\rm \micro\second$), the final absolute time accuracy is $\sqrt{2.19^2-2^2-0.1^2}=0.89\ \rm \micro\second$, which is consistent with 0 within 5.6 $\sigma$ error.

\subsection{Between GECAM-C and GECAM-B}
Crab pulsar have a negative lag between different energy bands (i.e. high-energy photons follow low-energy photons) with $t_{\rm lag}(E)=-A(E/{\rm keV})+C$ ($A\approx 0.27\ {\rm \mu s/keV}$) \citep{molkov2009absolute,tuo2022orbit}, which can be explained as having two independent and different spectral and phase distribution 
components \citep{massaro2000fine}. Since the detector design and energy response of GECAM-C and GBM are not identical, systematic error may be introduced in the time calibration. Therefore, we utilize the Crab pulsar data observed by GECAM-C and the very similar GECAM-B for time calibration. The time difference between the pulsar profiles observed by GECAM-C and GECAM-B in each month and the results are shown in Table \ref{allresults}; the final absolute time accuracy is -5.82$\pm$ 4.11 $\rm \micro\second$ (reduced-$\chi^2$=0.86) through fitting (see the bottom panel in Figure \ref{errors}). Therefore, there is no significant difference in the time difference between GECAM and GBM versus GECAM-C and GECAM-B. If we do not take into account the error introduced by the time resolution of GECAM-C (i.e. 0.1 $\rm \micro\second$) and GECAM-B (i.e. 0.1 $\rm \micro\second$), the final absolute time accuracy is $\sqrt{4.11^2-0.1^2-0.1^2}=3.59\ \rm \micro\second$, which is consistent with 0 within 1.6 $\sigma$ error.

In addition, we can calculate the time difference between GECAM-B and GECAM-C by using the time difference between GECAM-C and GBM, and between GECAM-B and GBM, and the error is obtained by error transfer. The result is $-2.96\pm2.42\ \rm \micro\second$, which is consistent with the time difference obtained directly calculation (i.e. $-5.82\pm3.59\ \rm \micro\second$) within the error.

\begin{center} 
\begin{table*}
\caption{\centering Time difference between Crab pulse profiles observed by GECAM-C and GBM, GECAM-C and GECAM-B, as well as GECAM-B and GBM in each month.}
\label{crab_table}
\begin{tabular*}{1\hsize}{@{}@{\extracolsep{\fill}}rccc@{}} 
	\hline
	\makecell[c]{Time\\(month)} & \makecell[c]{Time difference Between\\GECAM-C and GBM ($\rm \micro\second$)}& \makecell[c]{Time difference Between\\GECAM-C and GECAM-B ($\rm \micro\second$)}&\makecell[c]{Time difference Between\\GECAM-B and GBM ($\rm \micro\second$)}\\
	\hline
202208 & $22.30\pm14.36$ & $-24.33\pm22.43$ & $8.11\pm7.94$ \\
202209 & $-22.30\pm11.43$ & $36.49\pm22.09$ & $-4.05\pm13.60$ \\
202210 & $2.03\pm14.79$ & $-20.27\pm26.71$ & $8.11\pm19.02$ \\
202211 & $10.14\pm17.70$ & $-18.25\pm15.05$ & $4.05\pm7.66$ \\
202212 & $-4.05\pm9.63$ & $0.00\pm10.96$ & $6.08\pm5.96$ \\
202301 & $2.03\pm5.75$ & $6.08\pm12.42$ & $-12.16\pm9.95$ \\
202302 & $14.19\pm7.05$ & $-24.33\pm16.22$ & $14.19\pm12.31$ \\
202303 & $-4.05\pm12.33$ & $8.11\pm14.05$ & $-4.05\pm8.52$ \\
202304 & $-6.08\pm9.01$ & $-16.22\pm13.24$ & $18.25\pm8.34$ \\
202305 & $8.11\pm13.23$ & $-4.05\pm16.93$ & $6.08\pm9.69$ \\
202306 & $8.11\pm11.57$ & $-12.16\pm16.86$ & $8.11\pm8.00$ \\
202307 & $-6.08\pm11.76$ & $-10.14\pm13.11$ & $2.03\pm8.72$ \\
	\hline
\end{tabular*}
\label{allresults}
\end{table*}
\end{center}

\begin{table}
	\caption{\centering Summary of the time calibration results of GECAM.}
	\label{CONCLU}
	\begin{tabular}{lcr} 
		\hline
		 Detectors &  Time accuracy (1$\sigma$)\\
		\hline
		Detectors on GECAM-C (time resolution) & 0.1 $\rm \micro\second$\\
		GECAM-C and GBM & 2.02$\pm2.26$ $\rm \micro\second$\\
		GECAM-C and GECAM-B & -5.82$\pm3.59$ $\rm \micro\second$\\
		GECAM-B and GBM & 4.98$\pm0.89$ $\rm \micro\second$\\
		\hline
	\end{tabular}
 \label{zongde}
\end{table}

\section{Conclusion}
In this work, we performed the relative time calibration of the detectors on board GECAM-C and the absolute time calibration using the one-year observational data from its launch in 2022 Aug to 2023 July. The results are summarized in Table \ref{zongde}.

The relative time accuracy of the detector on board GECAM-C is calibrated to $\sim$ 0.1 $\rm \micro\second$ (1$\sigma$) using observed secondary particles produced almost simultaneously by cosmic rays hitting the satellite, which is consistent with the result of GECAM-B. The time accuracy between detectors in two DAQs is slightly worse than that of detectors in the same DAQ. Besides, we further verify the accuracy is stable by analyzing data in each month, which also indicates that GECAM-C, as same as GECAM-B, has the highest time resolution among all GRB missions ever flown, such as 2 $\rm \micro\second$ for Fermi/GBM and HXMT/HE (\citealp{2009ApJ...702..791M};\citealp{liu2020high}), 10 $\rm \micro\second$ for AstroSAT \citep{agrawal2006broad}, 100 $\rm \micro\second$ for {\it Swift}/BAT \citep{sakamoto2008first}, 2 ms for Konus-Wind \citep{aptekar1995konus} and 50 ms for SPI-ACS \citep{von2003integral}.

The absolute time accuracy of GECAM-C is calibrated to $2.02\pm 2.26\ \mu \rm s$ by the one-year Crab pulsar data observed jointly with Fermi/GBM, and we also investigate the accuracy in each month and confirm that it is stable (consistent within the error). Besides, to investigate the time-calibration systematic errors that may result from the Crab pulsar spectral lags in different energies (since different instruments have different energy responses), we calibrated the time accuracy between GECAM-C and GECAM-B, as well as between GECAM-B and GBM, and the results are $-5.82\pm 3.59\ \mu \rm s$ and $4.98\pm 0.89\ \mu \rm s$, respectively.
Since GECAM and GBM have similar energy responses, there is no significant difference in the time difference between GECAM and GBM versus GECAM-C and GECAM-B. However, the time difference between GECAM-B and GBM is consistent with 0 within 5.6 $\sigma$ error, and the difference in energy response may be one of the reasons.
If we consider the error introduced by the time resolution, the time difference between GECAM-C and GBM, GECAM-C and GECAM-B, as well as GECAM-B and GBM are $2.02\pm 3.02\ \mu \rm s$, $-5.82\pm 4.11\ \mu \rm s$ and $4.98\pm 2.19\ \mu \rm s$, respectively, which are all consistent with 0 within 3 $\sigma$.

For the high time resolution and absolute time accuracy, GECAM not only can identify cosmic ray events with very few false positives, but also has advantages as well as potential in timing analysis, for example, \\
(1) Time delay localization is a well-known method of localizing bursts by calculating the time difference between signals received by satellites at different positions (i.e. requires high time resolution and absolute time accuracy) (e.g. \citealp{hurley2011interplanetary,pal2013interplanetary,xiao2021enhanced}), and the precise location of GRB and soft gamma repeater (SGR) is crucial in multimessenger and multiwavelength astronomy for guiding follow-up observational, as well as identifying associated transients, neutrinos and gravitational waves. The absolute time accuracy 3 $\mu \rm s$ corresponds to only about 0.01 degrees in the time delay localization in low Earth orbits satellites. More importantly, the high time resolution of 0.1 $\mu \rm s$ combined with the Li-CCF method can give more accurate locations \citep{li2004timescale,xiao2021enhanced}.\\
(2) Timing analysis is an important approach to revealing radiation mechanisms and  progenitors, such as spectral lag between different energy bands (e.g. \citealp{zhang2009discerning,ukwatta2012lag,bernardini2014comparing}) and minimum variation timescale (i.e. requires high time resolution) (e.g. \citealp{1978Natur.271..525S, fenimore1993escape, 2007ApJ...660..556T,ackermann2014fermi, golkhou2014uncovering,golkhou2015energy}). Meanwhile, in the future, the sensitivity of the detector should be considered to be improved to obtain signals with higher signal-to-noise ratios in order to best utilize the advantages of high time resolution.\\
(3) The quasiperiodic oscillation (i.e. requires high time resolution for high-frequency searches) of magnetar (e.g. \citealp{huppenkothen2014quasi,xiao2022quasi,xiao2022search,xiao2023individual,roberts2023quasi}) can be used to constrain both the equation of state and the interior magnetic field of the neutron star via asteroseismology. Besides, it is worth noting that the impact of dead time on the search for high-frequency QPOs are significant, and thus further decreasing the dead time should be considered. \\
(4) Cosmic rays hit a detector or the
satellite structure can simultaneously produce many secondary
particles recorded by multiple GRDs and CPDs nearly
simultaneously (i.e. requires high relative time accuracy to identify); for the time resolution 0.1 $\mu \rm s$, both detectors have only $\sim$ 0.2 false positives per second if they both have a count rate of 1000. Besides, relative time accuracy is also critical when performing joint analysis using multiple detectors. \\
(5) Pulsar navigation experiment is important for both orbit estimation of Earth satellites and deep space navigation for spacecraft (e.g. \citealp{witze2018nasa,zheng2019orbit,luo2023pulsar}), which needs that the arrival times of events observed are corrected to the solar system barycenter (DE200) (i.e. requires high time resolution and absolute time accuracy); the absolute time accuracy of 3 $\mu \rm s$ corresponds to a light travel distance of only about 900 meters.\\

Finally, an interesting by-product of this work is the spectral lag between different energy bands of the Crab pulsar observed by GECAM and GBM. As shown in Figure \ref{lag_ee}, the energy and spectral lag relations of the Crab pulsar observed by GECAM-C, GECAM-B, and GBM can be fitted with a linear function with slopes $-0.17\pm0.10\ \mu\second\ \rm keV^{-1}$, $-0.20\pm0.12\ \mu\second\ \rm keV^{-1}$ and $-0.19\pm0.03\ \mu\second\ \rm keV^{-1}$, respectively. These results are consistent within 3 $\sigma$ error with those previously obtained through RXTE/PCA, HEXTE, and INTEGRAL/
SPI ($-0.6\pm0.2\ \mu\second\ \rm keV^{-1}$), as well as NICER and HXMT ($-0.26\pm0.02\ \mu\second\ \rm keV^{-1}$), however, the rates of decrease we measured are slightly smaller.

\section*{Acknowledgments}
We acknowledge the public data from {\it Fermi}/GBM. This work is supported by the National Key R\&D Program of China 2022YFF0711404, the Scientific Research Project of the Guizhou Provincial Education (Nos. KY[2022]123, KY[2022]132, KY[2022]137, KY[2021]303, KY[2020]003 and KY[2023]059).
The authors also thank supports from 
the Strategic Priority Research Program on Space Science, the Chinese Academy of Sciences (Grant No.
XDA15010100, 
XDA15360100, XDA15360102, 
XDA15360300, 
XDA15052700), 
the National Natural Science Foundation of China (Projects: 12273042, 12061131007, Grant No. 12173038, Nos. 12273008 and 12103013), the Foundation of Education Bureau of Guizhou Province, China (Grant No. KY (2020) 003), Guizhou Provincial Science and Technology Foundation (Nos. ZK[2022]304, [2021]023, [2023]024).

\newpage

\bibliography{main}

\begin{thebibliography}{}
\expandafter\ifx\csname natexlab\endcsname\relax\def\natexlab#1{#1}\fi
\providecommand{\url}[1]{\href{#1}{#1}}
\providecommand{\dodoi}[1]{doi:~\href{http://doi.org/#1}{\nolinkurl{#1}}}
\providecommand{\doeprint}[1]{\href{http://ascl.net/#1}{\nolinkurl{http://ascl.net/#1}}}
\providecommand{\doarXiv}[1]{\href{https://arxiv.org/abs/#1}{\nolinkurl{https://arxiv.org/abs/#1}}}

\bibitem[{Ackermann {et~al.}(2014)Ackermann, Ajello, Asano, Atwood, Axelsson,
  Baldini, Ballet, Barbiellini, Baring, Bastieri,
  {et~al.}}]{ackermann2014fermi}
Ackermann, M., Ajello, M., Asano, K., {et~al.} 2014, Science, 343, 42

\bibitem[{Agrawal(2006)}]{agrawal2006broad}
Agrawal, P. 2006, Advances in Space Research, 38, 2989

\bibitem[{An {et~al.}(2023)An, Antier, Bi, Bu, Cai, Cao, Camisasca, Chang,
  Chen, Chen, {et~al.}}]{an2023insight}
An, Z.-H., Antier, S., Bi, X.-Z., {et~al.} 2023, arXiv preprint
  arXiv:2303.01203

\bibitem[{Aptekar {et~al.}(1995)Aptekar, Frederiks, Golenetskii, Ilynskii,
  Mazets, Panov, Sokolova, Terekhov, Sheshin, Cline,
  {et~al.}}]{aptekar1995konus}
Aptekar, R., Frederiks, D., Golenetskii, S., {et~al.} 1995, Space Science
  Reviews, 71, 265

\bibitem[{Bachetti {et~al.}(2021)Bachetti, Markwardt, Grefenstette, Gotthelf,
  Kuiper, Barret, Cook, Davis, F{\"u}rst, Forster,
  {et~al.}}]{bachetti2021timing}
Bachetti, M., Markwardt, C.~B., Grefenstette, B.~W., {et~al.} 2021, The
  Astrophysical Journal, 908, 184

\bibitem[{Bak {et~al.}(1987)Bak, Tang, \& Wiesenfeld}]{bak1987self}
Bak, P., Tang, C., \& Wiesenfeld, K. 1987, Physical review letters, 59, 381

\bibitem[{Basu {et~al.}(2018)Basu, Joshi, Bhattacharya, Rao, Naidu,
  Krishnakumar, Arumugsamy, Vadawale, Manoharan, Dewangan,
  {et~al.}}]{basu2018timing}
Basu, A., Joshi, B.~C., Bhattacharya, D., {et~al.} 2018, Astronomy \&
  Astrophysics, 617, A22

\bibitem[{Bernardini {et~al.}(2014)Bernardini, Ghirlanda, Campana, Covino,
  Salvaterra, Atteia, Burlon, Calderone, D'Avanzo, D'Elia,
  {et~al.}}]{bernardini2014comparing}
Bernardini, M.~G., Ghirlanda, G., Campana, S., {et~al.} 2014, Monthly Notices
  of the Royal Astronomical Society, 446, 1129

\bibitem[{Chen {et~al.}(2021)Chen, Xiao, Xiong, Yu, Wen, Gong, Li, Li, Hou,
  Yang, {et~al.}}]{chen2021design}
Chen, C., Xiao, S., Xiong, S., {et~al.} 2021, Experimental Astronomy, 1

\bibitem[{Cusumano {et~al.}(2012)Cusumano, La~Parola, Capalbi, Perri,
  Beardmore, Burrows, Campana, Kennea, Osborne, Sbarufatti,
  {et~al.}}]{cusumano2012timing}
Cusumano, G., La~Parola, V., Capalbi, M., {et~al.} 2012, Astronomy \&
  Astrophysics, 548, A28

\bibitem[{Edwards {et~al.}(2006)Edwards, Hobbs, \&
  Manchester}]{edwards2006tempo2}
Edwards, R.~T., Hobbs, G., \& Manchester, R. 2006, Monthly Notices of the Royal
  Astronomical Society, 372, 1549

\bibitem[{Fenimore {et~al.}(1993)Fenimore, Epstein, \& Ho}]{fenimore1993escape}
Fenimore, E., Epstein, R., \& Ho, C. 1993, Astronomy and Astrophysics
  Supplement Series, 97, 59

\bibitem[{Golkhou \& Butler(2014)}]{golkhou2014uncovering}
Golkhou, V.~Z., \& Butler, N.~R. 2014, The Astrophysical Journal, 787, 90

\bibitem[{Golkhou {et~al.}(2015)Golkhou, Butler, \&
  Littlejohns}]{golkhou2015energy}
Golkhou, V.~Z., Butler, N.~R., \& Littlejohns, O.~M. 2015, The Astrophysical
  Journal, 811, 93

\bibitem[{Huppenkothen {et~al.}(2014)Huppenkothen, Heil, Watts, \&
  G{\"o}{\u{g}}{\"u}{\c{s}}}]{huppenkothen2014quasi}
Huppenkothen, D., Heil, L., Watts, A., \& G{\"o}{\u{g}}{\"u}{\c{s}}, E. 2014,
  The Astrophysical Journal, 795, 114

\bibitem[{Hurley {et~al.}(2011)Hurley, Briggs, Kippen, Kouveliotou, Fishman,
  Meegan, Cline, Trombka, McClanahan, Boynton,
  {et~al.}}]{hurley2011interplanetary}
Hurley, K., Briggs, M., Kippen, R., {et~al.} 2011, The Astrophysical Journal
  Supplement Series, 196, 1

\bibitem[{Kirsch {et~al.}(2004)Kirsch, Becker, Benlloch-Garcia, Jansen,
  Kendziorra, Kuster, Lammers, Pollock, Possanzini, Serpell,
  {et~al.}}]{kirsch2004timing}
Kirsch, M.~G., Becker, W., Benlloch-Garcia, S., {et~al.} 2004, in X-Ray and
  Gamma-Ray Instrumentation for Astronomy XIII, Vol. 5165, International
  Society for Optics and Photonics, 85--95

\bibitem[{{Kuiper} {et~al.}(2003){Kuiper}, {Hermsen}, {Walter}, \&
  {Foschini}}]{2003A&A...411L..31K}
{Kuiper}, L., {Hermsen}, W., {Walter}, R., \& {Foschini}, L. 2003, \aap, 411,
  L31, \dodoi{10.1051/0004-6361:20031353}

\bibitem[{Li {et~al.}(2004)Li, Qu, Feng, Song, Ding, \& Chen}]{li2004timescale}
Li, T.-P., Qu, J.-L., Feng, H., {et~al.} 2004, Chinese Journal of Astronomy and
  Astrophysics, 4, 583

\bibitem[{Li {et~al.}(2021)Li, Wen, Xiong, Gong, Zhang, An, Xu, Liu, Cai,
  Chang, {et~al.}}]{li2021inflight}
Li, X., Wen, X., Xiong, S., {et~al.} 2021, arXiv preprint arXiv:2112.04772

\bibitem[{Liu {et~al.}(2020)Liu, Zhang, Li, Lu, Chang, Li, Zhang, Jin, Yu,
  Zhang, {et~al.}}]{liu2020high}
Liu, C., Zhang, Y., Li, X., {et~al.} 2020, SCIENCE CHINA Physics, Mechanics \&
  Astronomy, 63, 1

\bibitem[{Luo {et~al.}(2023)Luo, Xiao, Zheng, Ge, Tuo, Xiong, Zhang, Lu, Huang,
  Yang, {et~al.}}]{luo2023pulsar}
Luo, X.-H., Xiao, S., Zheng, S.-J., {et~al.} 2023, The Astrophysical Journal
  Supplement Series, 266, 16

\bibitem[{Lyne {et~al.}(1993)Lyne, Pritchard, \& Graham~Smith}]{lyne199323}
Lyne, A., Pritchard, R., \& Graham~Smith, F. 1993, Monthly Notices of the Royal
  Astronomical Society, 265, 1003

\bibitem[{Massaro {et~al.}(2000)Massaro, Cusumano, Litterio, \&
  Mineo}]{massaro2000fine}
Massaro, E., Cusumano, G., Litterio, M., \& Mineo, T. 2000, arXiv preprint
  astro-ph/0006064

\bibitem[{{Meegan} {et~al.}(2009){Meegan}, {Lichti}, {Bhat}, {Bissaldi},
  {Briggs}, {Connaughton}, {Diehl}, {Fishman}, {Greiner}, {Hoover}, {van der
  Horst}, {von Kienlin}, {Kippen}, {Kouveliotou}, {McBreen}, {Paciesas},
  {Preece}, {Steinle}, {Wallace}, {Wilson}, \&
  {Wilson-Hodge}}]{2009ApJ...702..791M}
{Meegan}, C., {Lichti}, G., {Bhat}, P.~N., {et~al.} 2009, \apj, 702, 791,
  \dodoi{10.1088/0004-637X/702/1/791}

\bibitem[{Molkov {et~al.}(2009)Molkov, Jourdain, \&
  Roques}]{molkov2009absolute}
Molkov, S., Jourdain, E., \& Roques, J. 2009, The Astrophysical Journal, 708,
  403

\bibitem[{Paciesas {et~al.}(2012)Paciesas, Meegan, von Kienlin, Bhat, Bissaldi,
  Briggs, Burgess, Chaplin, Connaughton, Diehl, {et~al.}}]{paciesas2012fermi}
Paciesas, W.~S., Meegan, C.~A., von Kienlin, A., {et~al.} 2012, The
  Astrophysical Journal Supplement Series, 199, 18

\bibitem[{Pal'Shin {et~al.}(2013)Pal'Shin, Hurley, Svinkin, Aptekar,
  Golenetskii, Frederiks, Mazets, Oleynik, Ulanov, Cline,
  {et~al.}}]{pal2013interplanetary}
Pal'Shin, V., Hurley, K., Svinkin, D., {et~al.} 2013, The Astrophysical Journal
  Supplement Series, 207, 38

\bibitem[{Roberts {et~al.}(2023)Roberts, Baring, Huppenkothen, Gogus, Kaneko,
  Kouveliotou, Lin, van~der Horst, \& Younes}]{roberts2023quasi}
Roberts, O.~J., Baring, M.~G., Huppenkothen, D., {et~al.} 2023, arXiv preprint
  arXiv:2306.08130

\bibitem[{Rots {et~al.}(2004)Rots, Jahoda, \& Lyne}]{rots2004absolute}
Rots, A.~H., Jahoda, K., \& Lyne, A.~G. 2004, The Astrophysical Journal
  Letters, 605, L129

\bibitem[{Sakamoto {et~al.}(2008)Sakamoto, Barthelmy, Barbier, Cummings,
  Fenimore, Gehrels, Hullinger, Krimm, Markwardt, Palmer,
  {et~al.}}]{sakamoto2008first}
Sakamoto, T., Barthelmy, S., Barbier, L., {et~al.} 2008, The Astrophysical
  Journal Supplement Series, 175, 179

\bibitem[{{Schmidt}(1978)}]{1978Natur.271..525S}
{Schmidt}, W.~K.~H. 1978, \nat, 271, 525, \dodoi{10.1038/271525a0}

\bibitem[{Terada {et~al.}(2008)Terada, Enoto, Miyawaki, Ishisaki, Dotani,
  Ebisawa, Ozaki, Ueda, Kuiper, Endo, {et~al.}}]{terada2008orbit}
Terada, Y., Enoto, T., Miyawaki, R., {et~al.} 2008, Publications of the
  Astronomical Society of Japan, 60, S25

\bibitem[{{Titarchuk} {et~al.}(2007){Titarchuk}, {Shaposhnikov}, \&
  {Arefiev}}]{2007ApJ...660..556T}
{Titarchuk}, L., {Shaposhnikov}, N., \& {Arefiev}, V. 2007, \apj, 660, 556,
  \dodoi{10.1086/512027}

\bibitem[{Tuo {et~al.}(2022)Tuo, Li, Ge, Nie, Song, Xu, Zheng, Lu, Zhang, Liu,
  {et~al.}}]{tuo2022orbit}
Tuo, Y., Li, X., Ge, M., {et~al.} 2022, The Astrophysical Journal Supplement
  Series, 259, 14

\bibitem[{Ukwatta {et~al.}(2012)Ukwatta, Dhuga, Stamatikos, Dermer, Sakamoto,
  Sonbas, Parke, Maximon, Linnemann, Bhat, {et~al.}}]{ukwatta2012lag}
Ukwatta, T., Dhuga, K., Stamatikos, M., {et~al.} 2012, Monthly Notices of the
  Royal Astronomical Society, 419, 614

\bibitem[{von Kienlin {et~al.}(2003)von Kienlin, Beckmann, Rau, Arend, Bennett,
  McBreen, Connell, Deluit, Hanlon, Hurley, {et~al.}}]{von2003integral}
von Kienlin, A., Beckmann, V., Rau, A., {et~al.} 2003, Astronomy \&
  Astrophysics, 411, L299

\bibitem[{Witze(2018)}]{witze2018nasa}
Witze, A. 2018, Nature, 553, 261

\bibitem[{Xiao {et~al.}(2021)Xiao, Xiong, Zhang, Song, Lu, Huang, Cai, Yi,
  Song, Chen, {et~al.}}]{xiao2021enhanced}
Xiao, S., Xiong, S., Zhang, S., {et~al.} 2021, The Astrophysical Journal, 920,
  43

\bibitem[{Xiao {et~al.}(2022{\natexlab{a}})Xiao, Liu, Peng, An, Xiong, Tuo,
  Gong, Zhang, Zhang, Zheng, {et~al.}}]{xiao2022ground}
Xiao, S., Liu, Y., Peng, W., {et~al.} 2022{\natexlab{a}}, Monthly Notices of
  the Royal Astronomical Society, 511, 964

\bibitem[{Xiao {et~al.}(2022{\natexlab{b}})Xiao, Xiong, Cai, Song, Zheng, Peng,
  Wang, Qiao, Guo, Wang, {et~al.}}]{xiao2022energetic}
Xiao, S., Xiong, S.-L., Cai, C., {et~al.} 2022{\natexlab{b}}, Monthly Notices
  of the Royal Astronomical Society, 514, 2397

\bibitem[{Xiao {et~al.}(2022{\natexlab{c}})Xiao, Xiong, Wang, Zhang, Gao,
  Zhang, Cai, Yi, Zhao, Tuo, {et~al.}}]{xiao2022robust}
Xiao, S., Xiong, S.-L., Wang, Y., {et~al.} 2022{\natexlab{c}}, The
  Astrophysical Journal Letters, 924, L29

\bibitem[{Xiao {et~al.}(2022{\natexlab{d}})Xiao, Zhang, Zhu, Xiong, Gao, Xu,
  Zhang, Peng, Li, Zhang, {et~al.}}]{xiao2022quasi}
Xiao, S., Zhang, Y.-Q., Zhu, Z.-P., {et~al.} 2022{\natexlab{d}}, arXiv preprint
  arXiv:2205.02186

\bibitem[{Xiao {et~al.}(2022{\natexlab{e}})Xiao, Peng, Zhang, Xiong, Li, Tuo,
  Gao, Wang, Xue, Zheng, {et~al.}}]{xiao2022search}
Xiao, S., Peng, W.-X., Zhang, S.-N., {et~al.} 2022{\natexlab{e}}, The
  Astrophysical Journal, 941, 166

\bibitem[{{Xiao} {et~al.}(2023{\natexlab{a}}){Xiao}, {Tuo}, {Zhang}, {Xiong},
  {Lin}, {Zhang}, {Wang}, {Xue}, {Cai}, {Gao}, {Li}, {Li}, {Zheng}, {Liu},
  {Wang}, {Wang}, {Peng}, {Liu}, {Li}, {Wen}, {An}, {Song}, {Zheng}, {Zhang},
  {Dong}, {Xie}, {Feng}, {Ma}, {Wang}, {Luo}, {Dang}, {Shang}, {Zhi}, \&
  {Li}}]{xiao2023discovery}
{Xiao}, S., {Tuo}, Y.-L., {Zhang}, S.-N., {et~al.} 2023{\natexlab{a}}, \mnras,
  521, 5308, \dodoi{10.1093/mnras/stad885}

\bibitem[{{Xiao} {et~al.}(2023{\natexlab{b}}){Xiao}, Yang, Luo, Xiong, Qu,
  Zhang, Xue, Li, Tuo, Dong, Zhao, Dang, Shang, Ma, Cai, Wang, Wang, Li, Yi,
  Zhang, Ge, Zheng, Song, Peng, Wen, Li, An, Xu, Wang, Zheng, Zhang, Liu,
  Zhang, Xie, Feng, Wang, \& Zhi}]{xiao2023minimum}
{Xiao}, S., Yang, J.-J., Luo, X.-H., {et~al.} 2023{\natexlab{b}}.
\newblock \doarXiv{2307.07079}

\bibitem[{{Xiao} {et~al.}(2023{\natexlab{c}}){Xiao}, Li, Xue, Xiong, Zhang,
  Peng, Dong, Tuo, Cai, Luo, {et~al.}}]{xiao2023individual}
{Xiao}, S., Li, X.-B., Xue, W.-C., {et~al.} 2023{\natexlab{c}}, arXiv preprint
  arXiv:2307.14884

\bibitem[{Xiong(2020)}]{xiong2020gecam}
Xiong, S. 2020, SCIENTIA SINICA Physica, Mechanica \& Astronomica, 50, 129501

\bibitem[{Xiong {et~al.}(2023)Xiong, Wang, \& Huang}]{xiong2023grb}
Xiong, S., Wang, C., \& Huang, Y. 2023, GRB Coordinates Network, 33406, 1

\bibitem[{Zhang {et~al.}(2009)Zhang, Zhang, Virgili, Liang, Kann, Wu, Proga,
  Lv, Toma, Meszaros, {et~al.}}]{zhang2009discerning}
Zhang, B., Zhang, B.-B., Virgili, F.~J., {et~al.} 2009, The Astrophysical
  Journal, 703, 1696

\bibitem[{Zhang {et~al.}(2019)Zhang, Li, Xiong, Li, Sun, An, Xu, Zhu, Peng,
  Wang, {et~al.}}]{zhang2019energy}
Zhang, D., Li, X., Xiong, S., {et~al.} 2019, Nuclear Instruments and Methods in
  Physics Research Section A: Accelerators, Spectrometers, Detectors and
  Associated Equipment, 921, 8

\bibitem[{Zhang {et~al.}(2023)Zhang, Zheng, Liu, An, Wang, Wen, Li, Sun, Gong,
  Liu, Liu, Yang, Peng, Qiao, Guo, Feng, Zhang, Xue, Tan, Cai, Xiao, Yi, Xu,
  Gao, Wang, Hou, Huang, Zhao, Ma, Wang, Wang, Li, Zhang, Zhang, Li, Wang,
  Liang, Wang, Li, Ye, Zheng, Song, Zhang, Chen, \& Xiong}]{ZHANG2023168586}
Zhang, D., Zheng, C., Liu, J., {et~al.} 2023, Nuclear Instruments and Methods
  in Physics Research Section A: Accelerators, Spectrometers, Detectors and
  Associated Equipment, 1056, 168586,
  \dodoi{https://doi.org/10.1016/j.nima.2023.168586}

\bibitem[{Zheng {et~al.}(2023)Zheng, An, Peng, Zhang, Xiong, Qiao, Zhang, Xue,
  Liu, Feng, {et~al.}}]{zheng2023ground}
Zheng, C., An, Z.-H., Peng, W.-X., {et~al.} 2023, arXiv preprint
  arXiv:2303.00687

\bibitem[{Zheng {et~al.}(2019)Zheng, Zhang, Lu, Wang, Gao, Li, Song, Ge, Han,
  Chen, {et~al.}}]{zheng2019orbit}
Zheng, S., Zhang, S., Lu, F., {et~al.} 2019, The Astrophysical Journal
  Supplement Series, 244, 1

\end{thebibliography}

\end{CJK}
\end{document}